\documentstyle [12pt,epsf] {article}

\catcode`@=12
\newcommand{\beq}{\begin{equation}}
\newcommand{\eeq}{\end{equation}}

\newcommand{\bea}{\begin{eqnarray}}
\newcommand{\eea}{\end{eqnarray}}

\begin{document}
\thispagestyle{empty}
\begin{flushright} UCRHEP-T281\\June 2000\
\end{flushright}
\vspace{0.5in}
\begin{center}
{\Large \bf Light Sterile Neutrinos from Large Extra Dimensions\\}
\vspace{1.5in}
{\bf Ernest Ma$^1$, G. Rajasekaran$^{1,2}$, and Utpal Sarkar$^{1,3}$\\}
\vspace{0.2in}
{$^1$ \sl Department of Physics, University of California, Riverside, 
California 92521, USA\\}
\vspace{0.1in}
{$^2$ \sl Institute of Mathematical Sciences, Madras 600 113, India\\}
\vspace{0.1in}
{$^3$ \sl Physical Research Laboratory, Ahmedabad 380 009, India\\}
\vspace{1.5in}
\end{center}
\begin{abstract}

An experimentally verifiable Higgs-triplet model of neutrino masses from 
large extra dimensions was recently proposed.  We extend it to accomodate 
a light sterile neutrino which also mixes with the three active neutrinos. 
A previously proposed phenomenological model of four neutrinos (\underline 
{the only viable such model now left}, in view of the latest atmospheric 
and solar neutrino-oscillation data) is specifically realized.
\end{abstract}

\newpage
\baselineskip 24pt

In the standard model of particle interactions, neutrinos $\nu_i$ belong in 
left-handed doublets $(\nu_i, l_i)$ under the electroweak $SU(2)_L \times 
U(1)_Y$ gauge group.  Without right-handed singlet partners, they are 
massless but could obtain small Majorana masses through the effective 
dimension-5 operator 
\cite{wein}
\begin{equation}
{f_{ij} \over \Lambda} (\nu_i \phi^0 - l_i \phi^+)(\nu_j \phi^0 - l_j 
\phi^+),
\end{equation}
as the Higgs doublet $(\phi^+, \phi^0)$ acquires a nonzero vacuum expectation 
value ($vev$).  The parameter $\Lambda$ is an effective large mass scale. 

Different models of neutrino mass are merely specific realizations \cite{ma} 
of this operator in different extensions of the standard model.  The usual 
approach is to identify the lepton-number violation with a very large scale, 
i.e. $\langle \phi^0 \rangle = v << \Lambda$.  However, it is also possible 
that $1/\Lambda$ is actually of the form $m/M^2$, so that the scale of 
lepton-number violation may be associated with $m$ which happens to be very 
small.  Hence $M$ does not have to be very large, in which case the origin 
of neutrino mass may be tested directly in future high-energy colliders.  
In a recently proposed model \cite{extrip} in the context of large extra 
dimensions \cite{extra}, exactly this very desirable feature is accomplished 
by adding \cite{triplet} a Higgs triplet $(\xi^{++}, \xi^+, \xi^0)$ to the 
standard model, as well as a scalar singlet $\chi$ which also exists 
in the bulk \cite{exnu1}.

We now extend this proposal to include a light sterile neutrino $\nu_s$.  
(This is to be distinguished from other proposals \cite{exnu1,exnu2} where 
singlet fermions exist in the bulk which pair up with $\nu_i$ to become 
massive Dirac particles.  We note that strong bounds on the fundamental 
scale in such theories have also been obtained 
\cite{bound}.) The totality of present experimental evidence for neutrino 
oscillations \cite{atm,solar,lsnd} calls for 1 sterile neutrino in addition 
to the 3 active ones.  There are already a number of phenomenological 
scenarios for fitting all such data, but the latest results from the 
Super-Kamiokande Collaboration rule out the $\nu_\mu \to \nu_s$ explanation 
of the atmospheric data \cite{sobel} at the 99\% confidence level, and also 
rule out the $\nu_e \to \nu_s$ explanation of the solar data \cite{suzuki} 
at the 95\% confidence level.  This means that $\nu_s$ must be used in 
explaining the LSND result \cite{lsnd,mills} because a third 
mass-squared difference is required, and as far as we know, there 
is only one such phenomenological model that has been previously proposed 
\cite{mrs}.  The salient feature of this model is the fast decay of $\nu_s$ 
into an antineutrino $\bar \nu_i$ and a massless Goldstone boson (the Majoron) 
corresponding to the spontaneous breaking of lepton number.  We show in the 
following how this may naturally occur in the context of large extra 
dimensions.

The scalar singlet $\chi$ carries lepton number $L = -2$ \cite{extrip} and 
its $vev$ is the source of all lepton-number violation in our four-dimensional 
world (called a 3-brane).  The smallness of $\langle \chi \rangle$ 
is due to the distance of our brane in the extra space dimensions from the 
source of the lepton-number violation located in another brane \cite{distant}. 
Let $y$ denote a point in the extra dimensions.  Our brane is located at 
$y=0$, whereas another 3-brane (the one providing the lepton-number 
violation) is at $y=y_*$.  We assume that they are separated by $|y_*|=r$, 
where $r$ is the radius of compactification of the extra space dimensions, 
which is only a few $\mu$m in magnitude. The fundamental scale $M_*$ in this 
theory is then related to the usual Planck scale $M_P = 2.4 \times 10^{18}$ 
GeV by the relation
\begin{equation}
r^n M_*^{n + 2} \sim M_P^2,
\end{equation}
where $n$ is the number of extra space dimensions.  The scalar singlet $\chi$ 
exists in the bulk and could thus communicate between the 2 branes.  For the 
source of lepton-number violation, one possibility is that a scalar field 
$\sigma$, carrying lepton number $L=2$, acquires a large $vev$ in the other 
brane.  Assuming all mass parameters are of order $M_*$, the field $\sigma$ 
has an interaction with $\chi$ given by
\begin{equation}
{\cal L} = \alpha M_*^2 \int d^4 x' \sigma(x') \chi (x', y=y_*),
\end{equation}
where $\alpha$ is a parameter of order unity. 

Once $\sigma$ acquires a $vev$, lepton number will be broken in the other 
brane.  This will then act as a point source in the extra space dimensions 
for our world, so that the profile of $\chi$ is given by the 
Yukawa potential in the transverse dimensions \cite{distant}:
\begin{equation}
\langle \chi (y=0)\rangle = \langle \sigma ( y = y_*) 
\rangle \Delta_n (r),
\end{equation}
where 
\begin{equation}
\Delta_n (r) = {1 \over (2 \pi )^{n \over 2}
M_*^{n- 2}} ~\left( {m_\chi \over r} \right)^{n-2 \over 2}
~K_{n - 2 \over 2} \left( m_\chi r \right),
\end{equation}
$K$ being the modified Bessel function.  Assuming that $\langle \sigma \rangle 
= M_*$, we then have the $shining$ of $\chi$ everywhere in our world 
corresponding to
\begin{equation}
\langle \chi \rangle \approx \displaystyle{
{ \Gamma ( {n -2 \over 2} ) \over
4 \pi^{n \over 2} }{M_* \over (M_* r)^{n-2} } } \approx \displaystyle{ 
{\Gamma ({n-2 \over 2}) \over 4 \pi^{n \over 2}}} M_* \left( {M_* \over M_P} 
\right)^{2 - {4 \over n}},
\end{equation}
where $n > 2$ and $m_\chi r << 1$ have also been assumed.  This $shined$ 
value of $\chi$ now appears as a boundary condition for our brane.  In other 
words, the localized fields in our world must interact with $\chi$ in such 
a way that $\langle \chi \rangle$ is unaffected.

Consider now the addition of a singlet left-handed sterile neutrino $\nu_s$. 
We propose the following 2 simple scenarios: (A) $\nu_s$ has $L = +1$, and 
(B) $\nu_s$ has $L = -1$.  In (A), the interaction Lagrangian involving 
$\nu_i$ and $\nu_s$ is given by
\begin{equation}
{\cal L} = f_{ij} [ \xi^0 \nu_i \nu_j + \xi^+ (\nu_i l_j
+ l_i \nu_j)/ \sqrt 2
 + \xi^{++} l_i l_j] + f_i \nu_s (\nu_i \eta^0
 - l_i \eta^+) + h.c.,
\end{equation}
where $(\xi^{++},\xi^+,\xi^0)$ is a scalar triplet and $(\eta^+,\eta^0)$ is 
a scalar doublet, both carrying $L = -2$.  We define $\langle \chi \rangle 
\equiv z$ and express the bulk field as
\begin{equation}
\chi = {1 \over \sqrt 2} ( \rho + z \sqrt 2) e^{i \varphi} .
\end{equation}
We assume that its behavior is not altered by the 
parameters in different branes.  All such effects are already included in 
the boundary condition of Eq.~(6).  The lepton-number conserving interactions 
of $\chi$ with the other scalar fields in our world are then contained in
\begin{eqnarray}
L &=& \int d^4 x \left[h z (y=0) 
e^{i \varphi(x)} \left(\bar \xi^0(x) \phi^0(x) \phi^0(x) - \sqrt 2 \xi^-(x) 
\phi^+(x) \phi^0 (x)+ \xi^{--}(x) \phi^+(x) \phi^+(x) \right) 
\right. \nonumber \\
&& \left. + ~\mu z (y=0) e^{i \varphi(x)}
\left(\bar \eta^0(x) \phi^0(x) + \eta^-(x) \phi^+(x) \right) 
+ h.c.  \right], \label{int}
\end{eqnarray}
where $\mu$ is a mass parameter which could be of order $M_*$ or the 
electroweak symmetry breaking scale. The self-interaction terms for the
bulk scalar are now given by
\begin{equation}
V(\chi) = \lambda z(y)^2 \rho(x,y)^2 + {1 \over \sqrt 2} \lambda
z(y) \rho(x,y)^3 + {1 \over 8} \lambda \rho(x,y)^4 .
\end{equation}
This formulation has the virtue of universality, i.e. $\lambda$ is unchanged, 
but $z$ can change depending on where our brane is from the distant brane. 
It is also invariant under $U(1)_L$ : $ \nu_i \to e^{i \theta} \nu_i$,  
$\nu_s \to e^{i \theta} \nu_s$, $\xi \to e^{-2 i \theta } \xi$,  $\eta \to 
e^{-2i\theta} \eta$, $\rho \to \rho$, and $\varphi \to \varphi - 2 \theta$. 
The form of the potential $V(\chi)$ is that of the usual spontaneously 
broken U(1) theory and is independent of parameters in our brane. 

The $shining$ of the field $\chi$ in our world induces a very weak 
lepton-number violating trilinear coupling of the Higgs triplet 
$\xi$ with the standard Higgs doublet $\Phi$ \cite{extrip}, 
as well as the mixing of the Higgs doublets $\eta$ and $\Phi$. In addition,
the $shined$ value of $\chi$ supplies a Majorana mass term to the sterile 
neutrino through the interaction
\begin{equation}
L = f_s \int d^4 x ~ z(y=0) e^{i\varphi(x)} \nu_s(x) \nu_s(x) + h.c.
\end{equation}
From the Lagrangian of Eq.~(9), it can easily be shown that $\xi^0$ and 
$\eta^0$ will have small $vev$s (say $u$ and $w$ respectively) which are 
proportional to $z$.  Although the $shined$ value of $\chi$, i.e. $z$, 
comes from lepton-number violation in a distant brane and may not be 
determined in terms of the parameters entering in our world, the $vev$s of 
the other fields will be obtained in the usual way by minimizing the 
appropriate Higgs potential.

Consider the following Higgs potential containing the fields $\xi$, $\eta$, 
and $\Phi$ with interaction terms involving $\chi$ in our brane:
\begin{eqnarray}
V &=& m^2 \Phi^\dagger \Phi + m_\xi^2 \xi^\dagger \xi 
+ m_\eta^2 \eta^\dagger \eta 
+ {1 \over 2} \lambda_1 (\Phi^\dagger \Phi)^2 + {1 \over 2} \lambda_2
(\xi^\dagger \xi)^2 + {1 \over 2} \lambda_3 (\eta^\dagger \eta)^2 
+ \lambda_4 (\Phi^\dagger \Phi)(\xi^\dagger \xi) \nonumber \\
&& + \lambda_5 (\Phi^\dagger \Phi)(\eta^\dagger \eta)
+ \lambda_6 (\xi^\dagger \xi)(\eta^\dagger \eta) 
 + \left( h z e^{i \varphi} \xi^\dagger \Phi \Phi + 
\mu z  e^{i \varphi} \eta^\dagger \Phi + h.c. \right)
\end{eqnarray}
Let $\langle \Phi \rangle = v$, $\langle \eta \rangle = w$ and 
$\langle \xi \rangle = u$, then the
conditions for the minimum of $V$ are given by
\begin{eqnarray}
v(m^2 + \lambda_1 v^2 + \lambda_4 u^2 + \lambda_5 w^2 
+ 2 h z u) + \mu z w &=& 0, \\
u(m_\xi^2 + \lambda_4 v^2 + \lambda_2 u^2 + \lambda_6 w^2) 
+ h v^2 z &=& 0, \\
w(m_\eta ^2 + \lambda_5 v^2 + \lambda_6 u^2 + \lambda_3 w^2) 
+ \mu v z &=& 0  .
\end{eqnarray}
These equations tell us that the $vev$s of the fields $\xi$
and $\eta$ are small, and are given by 
\begin{equation}
u \simeq - {h v^2 z \over m_\xi^2 + \lambda_4 v^2},  ~~~~~~ 
w \simeq - { \mu v z \over m_\eta^2 + \lambda_5 v^2},
\end{equation}
where $v \simeq \sqrt {-m^2/\lambda_1} \simeq 174$ GeV as usual. 
[In the above, we have neglected the term $\xi^\dagger \Phi \eta$ for 
simplicity.  Its presence would only change the expressions for $u$ 
and $w$, not their proportionality to $z$.  It also does not affect the 
composition of the Majoron in Eq.~(21) to be given below.] 
The $4 \times 4$ neutrino mass matrix including
the sterile neutrino in the basis $(\nu_i,\nu_s)$ is then of the form
\begin{equation}
{\cal M}_\nu = \pmatrix{ 2f_{ij}u & f_i w 
\cr f_i w & 2f_s z}, \label{mnu}
\end{equation}
where all the mass terms are naturally small, say of order 1 eV or less.

The kinetic-energy term of $\chi$ is
\begin{equation}
| \partial_\mu {1 \over \sqrt 2} ( \rho + z \sqrt 2) e^{i \varphi} |^2
= {1 \over 2} (\partial_\mu \rho)^2 + z^2  (\partial_\mu \varphi)^2 + ... ,
\end{equation}
which implies that $\sqrt 2 z \varphi$ is the properly normalized massless 
Goldstone boson (the Majoron) from the spontaneous breaking of lepton number 
in the bulk.  In the presence of the interaction terms of Eq.~(\ref{int}), 
other fields also participate in the spontaneous breaking of lepton number, 
hence the Majoron in our world becomes a combination of all these fields.  
The $4 \times 4$ mass matrix in the basis $({\rm Im} \phi^0,{\rm Im} \xi^0, 
{\rm Im} \eta^0, z \varphi)$ is given by
\begin{equation}
\pmatrix{ - 4 h z u - \mu z w / v & 2 h z v 
& \mu z & - 2 h u v - \mu w \cr 2 h z v 
& - h z v^2 / u & 0 & h v^2 \cr
\mu z & 0 & - \mu z v / w  &
\mu v \cr - 2 h u v - \mu w & h v^2 & \mu v & - h u v^2 / z - \mu w v / z } .
\end{equation}
Diagonalizing this matrix, we get one massless eigenstate
\begin{equation}
{v {\rm Im} \phi^0 + 2 u {\rm Im} \xi^0 + w {\rm Im} \eta^0 
\over \sqrt{ v^2 + 4 u^2 + w^2 }},
\end{equation}
which becomes the longitudinal component of the $Z$ boson, and another one 
which is the physical Majoron field:
\begin{equation}
N^{-{1 \over 2}} \left[ - v(w^2 + 2 u^2) {\rm Im} \phi^0 +
u (v^2 - w^2) {\rm Im} \xi^0 + w ( v^2 + 2 u^2) {\rm Im} \eta^0 
+ z(v^2 + w^2 + 4 u^2) z \varphi \right],
\end{equation}
where $N$ is a normalization constant. The Majoron
coupling matrix now becomes
\begin{equation}
{1 \over \sqrt{N}} \pmatrix{ 
2 f_{ij} u(v^2- w^2) & f_i w(v^2+ 2 u^2)
\cr f_i w(v^2+ 2 u^2) & 2 f_s z(v^2 + w^2 + 4 u^2)}
\end{equation}
In the limit $v \to \infty$, the Majoron coupling matrix and the neutrino 
mass matrix are simultaneously diagonalized, in which case the Majoron will 
have only diagonal couplings.  For finite $v$, the off-diagonal Majoron 
couplings to the neutrino mass eigenstates are suppressed by the factor 
$(w^2 + 2u^2)/v^2$, hence neutrino decay rates are very small and 
phenomenologically unimportant. 
 
In scenario (B), $\nu_s$ has $L = -1$.  Hence the scalar doublet $(\eta^+,
\eta^0)$ in Eq.~(7) now carries no lepton number, i.e. $L=0$.  To distinguish 
it from the usual Higgs doublet $(\phi^+,\phi^0)$, we add a discrete $Z_2$ 
symmetry, such that $\nu_s$ and $\eta$ are odd and all other fields are even. 
Instead of the $e^{i\varphi} \nu_s \nu_s$ term in Eq.~(11), we now have 
$e^{-i\varphi} \nu_s \nu_s$.  Instead of the $\mu z e^{i\varphi} \eta^\dagger 
\Phi$ term in Eq.~(12), we now have $\mu^2 \eta^\dagger \Phi$ which breaks 
the $Z_2$ discrete symmetry softly, and as such, the parameter $\mu$ can be 
small.  [We have neglected the term $h' z e^{i\varphi} \xi^\dagger \eta \eta$ 
for simplicity.  It does not affect the composition of the Majoron in Eq.~(25) 
to be given below.]

The equation of constraint for $u$ remains the same as Eq.(14), whereas 
the $\mu z w$ term in Eq.~(13) is replaced by $\mu^2 w$, and the $\mu v z$ 
term in Eq.~(15) is replaced by $\mu^2 v$.  Hence $u$ is the same as in 
Eq.~(16), and
\begin{equation}
w \simeq - {\mu^2 v \over m_\eta^2 + \lambda_5 v^2}.
\end{equation}
To obtain $w \sim 1$ eV, we need $\mu \sim 1$ MeV for $m_\eta \sim 1$ TeV.

The $4 \times 4$ mass matrix in the basis 
$({\rm Im} \Phi^0,{\rm Im} \xi^0,{\rm Im} \eta^0,z \varphi)$ now becomes
\begin{equation}
\pmatrix{ - 4 h z u - \mu^2 w / v & 2 h z v 
& \mu^2 & - 2 h u v - \mu w \cr 2 h z v 
& - h z v^2 / u & 0 & h v^2 \cr
\mu^2 & 0 & - \mu^2 v / w & 0 \cr 
- 2 h u v - \mu w & h v^2 & 0 & - h u v^2 / z } ,
\end{equation}
and the physical Majoron is
\begin{eqnarray}
&& N^{-{1 \over 2}} \left[ -  2 u^2 v {\rm Im} \Phi^0 +
u (v^2 + w^2) {\rm Im} \xi^0 - 2 u^2 w {\rm Im} \eta^0 + z(v^2 + w^2 + 4 u^2) 
z \varphi \right]  \nonumber \\
&& ~~~~~~~ \simeq (u {\rm Im} \xi^0 + z^2 \varphi) /  
\sqrt{u^2 + z^2} .  
\end{eqnarray}
Because $\nu_s$ has $L = -1$ instead of $L = +1$, the Majoron coupling matrix 
is no longer proportional to the neutrino mass matrix even in the limit of 
$v \to \infty$.  The latter remains the same as given by Eq.~(\ref{mnu}), but 
the former now differs from Eq.~(22), i.e.
\begin{equation}
{1 \over \sqrt {u^2 + z^2}} 
\pmatrix{ 2 f_{ij} u & 0
\cr 0 & - 2 f_s z} .
\end{equation}
When expressed in the basis of neutrino mass eigenstates, there are now 
unsuppressed off-diagonal terms.  Hence neutrino decay is fast and the 
phenomenological model of Ref.~\cite{mrs} is realized. 

We advocate thus the $4 \times 4$ neutrino mass matrix of Eq.~(17), where 
the diagonal $\nu_s$ entry is of order a few eV.  The $3 \times 3$ 
submatrix spanning the $\nu_e$, $\nu_\mu$, and $\nu_\tau$ neutrinos is 
used to accommodate the atmospheric \cite{atm,sobel} and solar 
\cite{solar,suzuki} neutrino data, and the mixing between $\nu_s$ and $\nu_i$ 
is used to fit the LSND data \cite{lsnd,mills}.  Fast decay of $\nu_s$ 
into $\bar \nu_i$ and the Majoron allows us to evade the constraints of the 
CDHSW accelerator experiment \cite{cdhsw}, as explained in Ref.~\cite{mrs}. 
This model can be tested in present and future solar-neutrino experiments by 
the observation of antineutrinos (instead of neutrinos) from the sun.  
Details are already given in Ref.~\cite{mrs}.

In conclusion, we have shown how a light sterile neutrino may be naturally 
accommodated in the context of a theory of large extra space dimensions 
where a scalar field $\chi$ in the bulk carries lepton number which is 
spontaneously broken in a distant brane.  The effects on our world are a small 
$vev$ for $\chi$ and its associated massless Goldstone boson (the Majoron). 
This allows us to have a naturally light $4 \times 4$ mass matrix including 
the 1 sterile and the 3 active neutrinos.  Two simple scenarios are discussed, 
one of which allows the fast decay of a heavier neutrino into a lighter 
antineutrino and the Majoron, as previously proposed \cite{mrs}.  In view 
of the latest experimental results which exclude $\nu_\mu \to \nu_s$ in 
atmospheric \cite{sobel} and $\nu_e \to \nu_s$ in solar \cite{suzuki} 
neutrino-oscillation data, this is the only viable such model now left.

{\it Acknowledgement.} This work was supported in part by the U.~S.~Department
of Energy under Grant No.~DE-FG03-94ER40837.  G.R. and U.S. also thank the 
UCR Physics Department for hospitality.

\bibliographystyle{unsrt}

\begin{thebibliography}{99}

\bibitem{wein} S. Weinberg, Phys. Rev. Lett. {\bf 43}, 1566 (1979).

\bibitem{ma} E. Ma, Phys. Rev. Lett. {\bf 81}, 1171 (1998).

\bibitem{extrip} E. Ma, M. Raidal and U. Sarkar, hep-ph/0006046.

\bibitem{extra} N. Arkani-Hamed, S. Dimopoulos, and G. Dvali, Phys. Lett.
{\bf B429}, 263 (1998);  I. Antoniadis, N. Arkani-Hamed, S. Dimopoulos,
and G. Dvali, Phys. Lett. {\bf B436}, 257 (1998); N. Arkani-Hamed, S.
Dimopoulos, and G. Dvali, Phys. Rev. {\bf D59}, 086004 (1999).

\bibitem{triplet} E. Ma and U. Sarkar, Phys. Rev. Lett. {\bf 80}, 5716 (1998).

\bibitem{exnu1} N. Arkani-Hamed, S. Dimopoulos, G. Dvali, and J. 
March-Russell, hep-ph/9811448. 

\bibitem{exnu2} K. R. Dienes, E. Dudas, and T. Gherghetta, Nucl. Phys. 
{\bf B557}, 25 (1999); A. E. Faraggi and M. Pospelov, Phys. Lett. {\bf B458}, 
237 (1999); A. Das and O. C. W. Kong, Phys. Lett. {\bf B470}, 149 (1999);
R. N. Mohapatra, S. Nandi, and A. Perez-Lorenzana, Phys. Lett. {B466}, 115 
(1999); R. N. Mohapatra and A. Perez-Lorenzana, hep-ph/9910474; 
hep-ph/0006278; A. Ioannisian and J. W. F. Valle, hep-ph/9911349;
G. Dvali and A. Yu. Smirnov, Nucl. Phys. {\bf B563}, 63 (1999); 
R. Barbieri, P. Creminelli, and A. Strumia, hep-ph/0002199.

\bibitem{bound}  A. Ioannisian and A. Pilaftsis, hep-ph/9907522;
L. N. Chang, O. Lebedev, W. Loinaz, and T. Takeuchi, hep-ph/0005236.

\bibitem{atm} Y. Fukuda {\it et al.}, Phys. Lett. {\bf B433}, 9 (1998); 
{\bf B436}, 33 (1998); Phys. Rev. Lett. {\bf 81}, 1562 (1998); {\bf 82}, 
2644 (1999).

\bibitem{solar} R. Davis, Prog. Part. Nucl. Phys. {\bf 32}, 13 (1994); P. 
Anselmann {\it et al.}, Phys. Lett. {\bf B357}, 237 (1995); {\bf B361}, 235 
(1996); J. N. Abdurashitov {\it et al.}, Phys. Lett. {\bf B328}, 234 (1994); 
Phys. Rev. Lett. {\bf 83}, 4686 (1999); Y. Fukuda {\it et al.}, Phys. Rev. 
Lett. {\bf 77}, 1683 (1996); {\bf 81}, 1158 (1998); {\bf 82}, 1810 (1999); 
{\bf 82}, 2430 (1999).

\bibitem{lsnd} C. Athanassopoulos {\it et al.}, Phys. Rev. Lett. {\bf 75}, 
2650 (1995); {\bf 77}, 3082 (1996); {\bf 81}, 1774 (1998).

\bibitem{sobel} H. Sobel, Talk given at {\it Neutrino 2000}, Sudbury, Canada 
(June 2000).

\bibitem{suzuki} Y. Suzuki, Talk given at {\it Neutrino 2000}, Sudbury, 
Canada (June 2000).

\bibitem{mills} G. Mills, Talk given at {\it Neutrino 2000}, Sudbury, Canada 
(June 2000).

\bibitem{mrs} E. Ma, G. Rajasekaran and I. Stancu, Phys. Rev. {\bf D61}, 
071302 (2000).

\bibitem{distant} N. Arkani-Hamed and S. Dimopoulos, hep-ph/9811353; N.
Arkani-Hamed, L. Hall, D. Smith, and N. Weiner, Phys. Rev. {\bf D61}, 116003
(2000); Y. Sakamura, hep-ph/9912511.

\bibitem{cdhsw} F. Dydak {\it et al.}, Phys. Lett. {\bf 134B}, 281 (1984).


\end{thebibliography}

\end{document}